\documentclass[fleqn,twoside]{article}                                         
\usepackage{espcrc2}
\usepackage{graphicx} 

\newcommand{\AmS}{{\protect\the\textfont2
  A\kern-.1667em\lower.5ex\hbox{M}\kern-.125emS}}
 
\newcommand{\ud}{\mathrm{d}}    
\title{Fast Calculations in Nonlinear Collective Models of Beam/Plasma Physics} 
\author{A.~N. Fedorova and M.~G. Zeitlin
\address{IPME, RAS, V.O. Bolshoj pr., 61,
 199178, St.~Petersburg, Russia\\
E-mail: zeitlin@math.ipme.ru, 
http://www.ipme.ru/zeitlin.html, 
http://www.ipme.nw.ru/zeitlin.html
}}

\begin{document}                                               

\thispagestyle{empty}

\begin{center}
\begin{tabular}{p{130mm}}

\begin{center}
{\bf\Large FAST CALCULATIONS IN NONLINEAR}\\
\vspace{5mm}

{\bf\Large COLLECTIVE MODELS OF} \\
\vspace{5mm}

{\bf\Large BEAM/PLASMA PHYSICS}\\

\vspace{1cm}

{\bf\Large Antonina N. Fedorova, Michael G. Zeitlin}\\

\vspace{1cm}

{\bf\it
IPME RAS, St.~Petersburg,
V.O. Bolshoj pr., 61, 199178, Russia}\\
{\bf\large\it e-mail: zeitlin@math.ipme.ru}\\
{\bf\large\it e-mail: anton@math.ipme.ru}\\
{\bf\large\it http://www.ipme.ru/zeitlin.html}\\
{\bf\large\it http://www.ipme.nw.ru/zeitlin.html}
\end{center}

\vspace{1cm}
\begin{center}
\begin{tabular}{p{100mm}}
We consider an application of va\-ri\-a\-ti\-o\-nal\--wa\-ve\-let 
approach to nonlinear collective
models of  beam/plasma physics: Vlasov/Bo\-l\-t\-z\-mann-like reduction 
from general BBGKY hierachy. 
We obtain fast convergent
multiresolution representations for solutions which allow to
consider
polynomial and rational type of nonlinearities. The solutions are
represented via the multiscale decomposition in nonlinear high-localized
eigenmodes (waveletons).
\end{tabular}
\end{center}
\vspace{40mm}

\begin{center}
{\large Presented at VIII International Workshop on}\\
{\large  Advanced Computing and Analysis Techniques in Physics Research,} \\
{\large Section III "Simulations and Computations in }\\
{\large Theoretical Physics and Phenomenology"}\\
{\large ACAT'2002, June 24-28, 2002, Moscow}
\end{center}
\end{tabular}
\end{center}
\newpage


\begin{abstract}
We consider an application of va\-ri\-a\-ti\-o\-nal\--wa\-ve\-let 
approach to nonlinear collective
models of  beam/plasma physics: Vlasov/Bo\-l\-t\-z\-mann-like reduction 
from general BBGKY hierachy. 
We obtain fast convergent
multiresolution representations for solutions which allow to
consider
polynomial and rational type of nonlinearities. The solutions are
represented via the multiscale decomposition in nonlinear high-localized
eigenmodes (waveletons).
\end{abstract}

\maketitle

We consider applications of numerical--ana\-ly\-ti\-cal technique based on 
variational-wavelet approach to nonlinear collective
models of  be\-am/\\
\noindent plas\-ma physics, e.g. some forms of Vla\-sov\-/Bol\-t\-zmann-like reductions 
from general BBGKY hierarchy.
These equations are related to the modeling of propagation of intense charged particle
beams in high-intensity accelerators and transport systems [1]. In our
approach we use fast convergent
multiresolution va\-ri\-a\-ti\-onal-wa\-ve\-let representations, which allows to
consider polynomial and rational type of nonlinearities [2], [3]. 
The solutions are
represented via the multiscale decomposition in nonlinear high-localized
eigenmodes, which corresponds to the full
multiresolution expansion in all underlying hidden time/space or
phase space scales. In contrast with different approaches we don't use
perturbation technique or linearization procedures. 
We consider representation (3) below, where 
each term corresponds to the contribution from the 
scale $i$ in the full underlying
multiresolution decomposition
as nonlinear multiscale
generalization of old  $\delta F$ approach [1].
As a result, fast scalar/parallel
modeling demonstrates appearance of high-localized coherent structures (waveletons)
and different pattern formation
in systems with complex collective behaviour.
Let M be the phase space of ensemble of N particles ($ {\rm dim}M=6N$)
with coordinates
$x_i=(q_i,p_i), \quad i=1,...,N, \quad
q_i=(q^1_i,q^2_i,q^3_i)\in R^3,\quad
p_i=(p^1_i,p^2_i,p^3_i)\in R^3$
and
$F_N(x_1,\dots,x_N;t)$
be the N-particle distribution functions. 
For $i=1,2$  we have from general BBGKY hierarchy:  
\begin{eqnarray}
&&\frac{\partial F_1(x_1;t)}{\partial t}+\frac{p_1}{m}\frac{\partial}{\partial q_1}
F_1(x_1;t)\\
&&=\int\ud x_2L_{12} F_2(x_1,x_2;t)\nonumber
\end{eqnarray}
\begin{eqnarray}
\frac{\partial F_2(x_1,x_2;t)}{\partial t}+\Big(\frac{p_1}{m}
\frac{\partial}{\partial q_1}+\frac{p_2}{m}\frac{\partial}{\partial q_2}-L_{12}\Big)
\end{eqnarray}
$$ F_2(x_1,x_2;t)
=\int\ud x_3(L_{13}+L_{23})F_3(x_1,x_2,x_3,t)
$$
where partial Liouvillean operators and details are described in [3].
We are interested in the cases when
$
F_k(x_1,\dots,x_k;t)=\prod^k_{i=1}F_1(x_i;t)+G_k(x_1,\dots,x_k;t)
$
where $G_k$ are correlators, really have additional reductions 
as in case of Vlasov-like systems.
Then we have in (1), (2) polynomial type of nonlinearities (more exactly, multilinearities).
Our goal is the demonstration of advantages of the following representation
\begin{equation}
F_j=\sum_{i\in Z}\delta^i F_j,
\end{equation}
for the full exact solution for the systems related to equations 
(1), (2). It is possible to consider (3) as multiscale 
nonlinear generalization of old $\delta F$ approach [1].
In (3) each $\delta^i F_j$ term corresponds to the contribution from the 
scale $i$ in the full underlying
multiresolution decomposition 
$
\dots\subset V_{-1}\subset V_0\subset V_1\subset V_2\dots
$
of the proper function space to which $F_j$ is really belong.
\begin{figure}[htb]                                                            
\centering                                                                      
\includegraphics*[width=60mm]{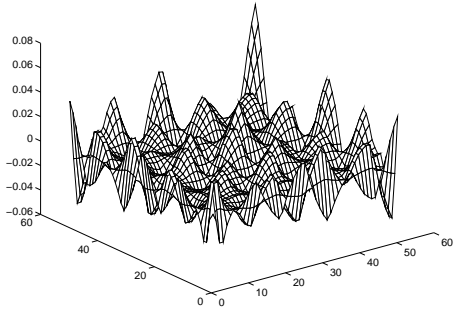}                              
\caption{$N=1$ eigenmode contribution to (3).}                                          
\end{figure}                  
It corresponds to the following decompositions:
\begin{eqnarray}
\{F_k(t)\}=\bigoplus_{-\infty<j<\infty} W_j,\  
\{F_k(t)\}=\overline{V_0\displaystyle\bigoplus^\infty_{j=0} W_j}
\end{eqnarray}
in case when $V_0$ is the coarsest scale of resolution and where 
$
V_{j+1}=V_j\bigoplus W_j
$
and bases in scale spaces $W_i(V_j)$ are generated from base functions $\psi(\varphi)$
by action of affine group of translations and dilations 
(the so called ``wavelet microscope'').
Our constructions are based on variational approach which provides 
the best possible
fast convergence properties
in the sense of combined norm  
$
\|F^{N+1}-F^{N}\|\leq\varepsilon
$
introduced in [3].
\begin{figure}[htb]                                                     
\centering                                                                      
\includegraphics*[width=60mm]{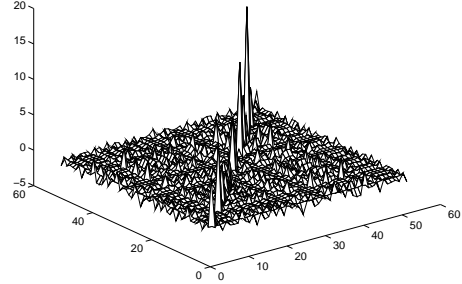}
\caption{Stable pattern/waveleton.}
\end{figure}
Our five basic points after functional space choice are:

\noindent
{\bf 1.} Ansatz-oriented choice of the (multi\-di\-men\-si\-o\-nal) ba\-ses
related to some po\-ly\-no\-mi\-al tensor algebra. Some examples related
to general BBGKY hierarchy are considered in [3].
\noindent
{\bf 2.} The choice of proper variational principle. A few 
pro\-je\-c\-ti\-on or Ga\-ler\-kin\--li\-ke 
principles for constructing (weak) solutions are considered in [2], [3].
It should be noted advantages of formulations related to biorthogonal (wavelet) decomposition.
\noindent
{\bf 3.} The choice of  bases functions in scale spaces $W_j$ from wavelet zoo. They 
correspond to high-localized (nonlinear) oscillations/excitations, 
coherent (nonlinear) resonances,
etc. Besides fast convergence properties of 
the corresponding va\-ri\-a\-ti\-o\-nal\--\-wa\-ve\-let expansions it should be noted 
minimal complexity of all underlying calculations, especially in case of choice of wavelet
packets which minimize Shannon entropy.
\noindent 
{\bf 4.} Operators  representations providing maximum sparse representations 
for arbitrary (pseu\-do)dif\-fe\-ren\-ti\-al/integral operators 
$\ud f/\ud x$, $\ud^n f/\ud x^n$, $\int T(x,y)f(y)\ud y)$, etc [3].
\noindent
{\bf 5.} (Multi)linearization. Besides variation appro\-ach we consider a different method
to deal with (po\-lynomial) nonlinearities, which is based on the para-product structures [3].

So, after application of points 1-5 above, we arrive to explicit 
numerical-analytical realization of
representations (3), (4). Fig.1 demonstrates the first contribution to
the full solution (3) while Fig.2 presents (stable) pattern (waveleton) as solution of system
(2), (3). We evaluate accuracy of calculations according to norm mentioned above [3].

 \end{document}